\documentclass[11pt,twoside]{article}
\usepackage{apn3conf}

\newcommand{\ea}{{\it et al.}}

\newcommand{\msol}{\mathrm{M}_\odot}

\newcommand{\kms}{km~$\rm{s}^{-1}$}

\newcommand{\beq}{\begin{equation}}
\newcommand{\eeq}{\end{equation}}
\newcommand{\bdm}{\begin{displaymath}}
\newcommand{\edm}{\end{displaymath}}

\begin{document}

\title{Clumpy Flows in Protoplanetary and Planetary Nebulae}

\author{Alexei Y. Poludnenko\altaffilmark{1} and Adam Frank\altaffilmark{2}}
\affil{Department of Physics and Astronomy, University of Rochester,
       Rochester, NY 14627-0171}
\author{Sorin Mitran\altaffilmark{3}}
\affil{Department of Mathematics, University of North Carolina,
       CB \#3250, Chapel Hill, NC 27599}

\altaffiltext{1}{wma@pas.rochester.edu}
\altaffiltext{2}{afrank@pas.rochester.edu}
\altaffiltext{3}{mitran@amath.unc.edu}

\contact{Alexei Y. Poludnenko}
\email{wma@pas.rochester.edu}

\paindex{Poludnenko, A. Y.}
\aindex{Frank, A.}
\aindex{Mitran, S.}

\authormark{Poludnenko, Frank, \& Mitran}


\begin{abstract}
  Many astrophysical flows occur in inhomogeneous media.  We briefly
  discuss some general properties of the adiabatic and radiative
  inhomogeneous systems and discuss the relevance of those properties
  to the planetary nebulae systems. We then focus on radiative
  hypersonic bullets and the applicability of this model to planetary
  and protoplanetary systems such as CRL 618, NGC 6543, Hen 3-1475.

\end{abstract}

\section{Introduction}
Recently it has been more and more widely acknowledged that it is
important to consider the inhomogeneous structure of stellar outflows
since "clumps", or "clouds", arising on a variety of scales can
introduce not only quantitative but also qualitative changes to the
overall dynamics of the flow. Planetary nebulae are an important
astrophysical example of flows where the presence of inhomogeneities
may play a conspicuous role in defining nebular dynamics and
morphology.

Embedded inhomogeneities may come in two flavors. One includes large
stationary or quasi-stationary ensembles of condensations, oftentimes
forming extended shells.  These may interact with the ambient flow,
e.g. cometary knot shells in NGC 7293 (Helix) or NGC 2392 (Eskimo).
The other includes more localized systems that consist of one or
several compact knots or ejecta moving at significant velocities
relative to the ambient medium. One of the most spectacular examples
of such nebular systems is CRL 618, which exhibits long thin shocked
lobes. The most prominent features of the lobes are their high
length-to-width ratio and the presence of the periodic rings in their
structure. There is also some evidence for a velocity increase in the
lobes from the base to the tip (S\'anchez Contreras, Sahai, \& Gil de
Paz 2002). Other examples include strings in NGC 6543 (Weis, Duschl,
\& Chu 1999) and knots in the outflow of Hen 3-1475 (Riera \ea 2003).

\section{Adiabatic vs. Radiative Systems}

When the interaction of an inhomogeneity with a global flow proceeds
in the adiabatic regime the dominant process is the \emph{lateral
  clump re-expansion} (Klein, McKee, \& Colella 1994; Poludnenko,
Frank, \& Blackman 2002). The internal shock compresses the clump
which then re-expands in the lateral direction.  Such re-expansion, in
combination with instabilities at the upstream clump surface, is
responsible for the destruction of the clump.

When more than one clump is present, lateral re-expansion also drives
interclump interactions via merging and formation of larger structures
that subsequently alter the global flow (Poludnenko \ea 2002). We
refer the reader to Poludnenko \ea (2002) for further details on the
dynamics of the adiabatic inhomogeneous systems. We note that two key
regimes can be defined: \emph{``interacting''} and
\emph{``noninteracting''} systems. Clumps in the interacting systems
will merge via lateral re-expansion before the clumps are destroyed.

Radiatively cooled inhomogeneous systems show a minimal role of
lateral re-expansion and, therefore, of interclump interaction. The
dominant process is the formation of instabilities with small
wavelength at the bullet upstream surface. Hydrodynamic instabilities
(Richtmeyer-Meshkov, Rayleigh-Taylor) produce the initial instability
seed. The resulting density variations quickly trigger the onset of
thermal instabilities which thereafter drive clump fragmentation.

The fragments produced by the instability are of different mass with
the most massive ones being closest to the symmetry axis and the
outermost ones being the lightest. As a result fragments are ``peeled
off'' from the clump starting with those outermost in radius. When
considering a single axisymmetric clump or bullet, each fragmentation
event results in the formation of a distinct \emph{``ring-like''
  feature} in the clump wake. The formation of rings is a consequence
of the axisymmetry. In 3-D it is likely that the rings would
themselves fragment (Klein \ea 2003).

\section{Radiative Bullet Simulations}

Here we present a simulation of the propagation of a hypersonic
radiative bullet as an example of a simple strongly cooled
inhomogeneous system. Since, as it was mentioned above, the evolution
of radiative inhomogeneous systems proceeds mostly in the
noninteracting regime, the case of a single bullet can be considered a
``building block'' of more complicated inhomogeneity configurations.
The simulation was performed with the AstroBEAR code based on the
BEARCLAW adaptive mesh refinement package.  Radiative cooling was
tracked via a standard cooling curve. We refer the reader to
(Poludnenko, Frank, \& Mitran 2003) for further details on the model
and the simulation.

The initial bullet radius is $r_b=3.0\cdot10^{15}$ cm, the initial
bullet density is $\rho_b=10^5 \ cm^{-3}$. This translates into the
initial bullet mass $m_b \approx 0.5\cdot10^{-5} \msol$. The initial
bullet Mach number is $M_b=20$, or $v_b \approx 235$ \kms.  The
computational domain is of size $1.8\cdot10^{17}$ cm $\times \ 
2.4\cdot10^{16}$ cm or expressed in terms of bullet radii $60r_b \ 
\times \ 8r_b$. The ambient density is $\rho_a=1000 \ cm^{-3}$ and the
ambient temperature $T_a=10^4 K$. The run was carried out in
cylindrical symmetry with the x-axis being the symmetry axis. We use
an adaptive grid with 4 levels of refinement with equivalent
resolution of $1024 \ \times \ 7680$ cells.

In the simulation the hydrodynamic timescale, i.e. the bullet crushing
time, was $t_{hydro} \ = \ 2.54\cdot10^9$ s $\approx \ 80.5$ yrs., the
cooling timescale was $t_{cool} \ = \ 70.9\cdot10^3$ s $\approx \ 
19.7$ hours. The simulation total run time was $9.49\cdot10^{9}$ s
$\approx 251.1$ yrs. Thus the evolution of the system was in a regime
strongly dominated by cooling.

\begin{figure}
\epsscale{1.0}
\plotone{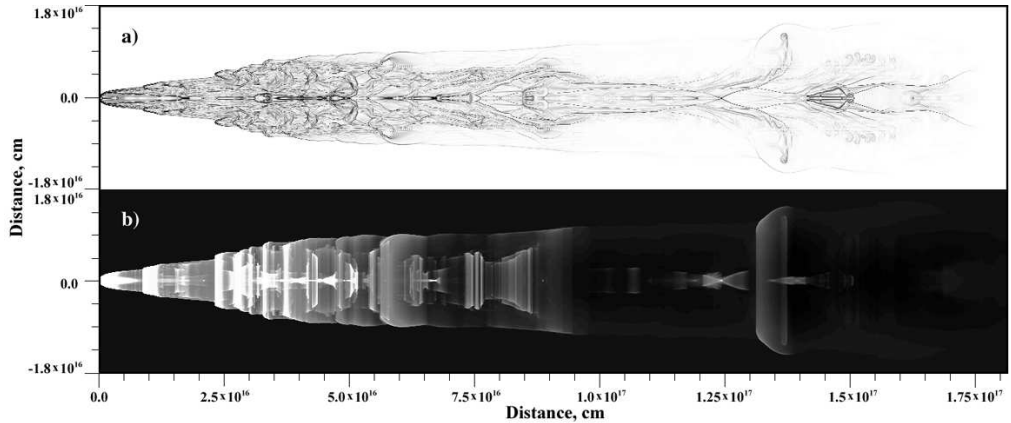}
\caption{a) Synthetic Schlieren image of the computational domain at time 251.1 yrs.
  Shown is the gradient of the density logarithm.  b) Synthetic
  observation image of the computational domain for the same time as
  in a).  Note the periodic ring-like structures in the domain
  resulting from individual fragmentation episodes.}
\label{LastFrame}
\end{figure}

Figure~\ref{LastFrame} shows the computational domain at time $t \ = \ 
251.1$ yrs. In Figure~\ref{LastFrame}a the synthetic Schlieren image
(gradient of the density logarithm) is shown illustrating the shock
and vortex sheet structure in the flow. Figure~\ref{LastFrame}b shows
the synthetic observation image of the computational domain. It
represents the 2D projected distribution of the logarithm of the
emissivity $I=n^2\Lambda$ integrated in the z-direction, where $n$ is
the number density and $\Lambda$ is the cooling function. Note the
presence of several characteristic ring-like structures in
Figure~\ref{LastFrame}b in the bullet wake, resulting from the
fragmentation events.

\begin{figure}
\epsscale{1.0}
\plotone{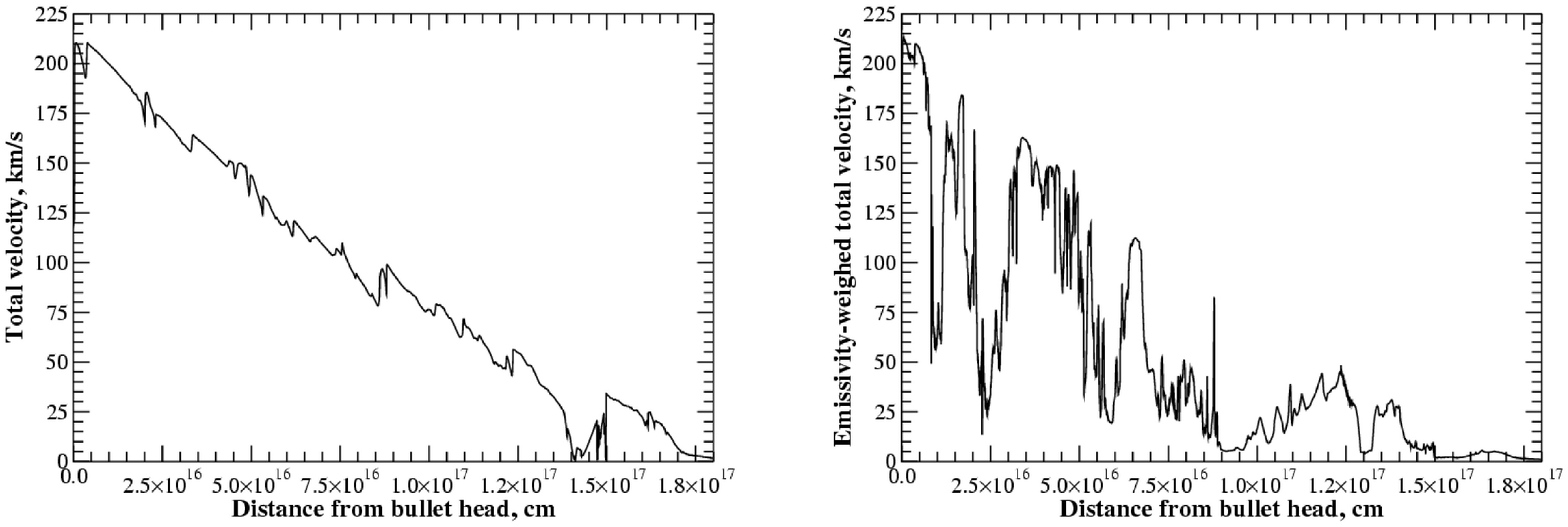}
\caption{\emph{Left:} distribution of total velocity along the symmetry axis of
  the bullet at the time 251.1 yrs. (same time in the simulation as
  the one shown in Figure~\ref{LastFrame}). \emph{Right:} distribution
  of the emissivity-weighed total velocity in the system at the same
  time.}
\label{Velocity}
\end{figure}

Another key property of such systems is the presence of Hubble-type
flows in the bullet wake. The left panel of Figure~\ref{Velocity}
shows the distribution of the total velocity $v_{tot}(x)$ along the
symmetry axis as a function of distance from the bullet head. The
linear velocity decrease from the maximum value of $210$ \kms from
head to base is clear aside from some minor fluctuations arising due
to the unsteadiness of the downstream flow. However, a better-founded
comparison with observational data needs quantities more closely
resembling observations. As an example, we consider the
emissivity-weighed total velocity $v_{emis}(x)$, which is shown in the
right panel of Figure~\ref{Velocity}. While it is significantly more
noisy than the total velocity, the local maxima roughly fall on the
same line as in the left panel giving some indication of a Hubble-type
flow. It should be emphasized that we found evidence of Hubble flows
only in systems dominated by radiative cooling, i.e.  where the ratio
of the cooling time scale over the hydrodynamic time scale is $\la 1$.

For radiative systems our principal conclusions are: (1) due to the
lack of lateral re-expansion such systems evolve mostly in the
noninteracting regime; (2) hypersonic radiative bullets are capable of
producing structures with high length-to-width ratios (between 6 and
10 for our study); (2) the dominant process responsible for bullet
destruction is instability formation at the bullet upstream interface
leading to separate fragmentation episodes which create periodic
``ring-like'' structures in the wake; (3) simulations of strongly
cooled systems show the presence of Hubble-type flows. How these flows
appear observationally remains an open question. These properties
suggest that hypersonic radiative bullets can be a likely candidate
for a model of such systems as shocked lobes of CRL 618, strings of
NGC 6543, etc.

This research was supported in part by the NSF grant AST-9702484, NASA
grant NAG5-8428, HST Grants, DOE and the Laboratory for Laser
Energetics under DOE sponsorship.

Animations can be found at
\htmladdURL{http://www.pas.rochester.edu/~wma}.

\end{document}